\documentstyle[aps,twocolumn,epsfig]{revtex}

\begin{document}
\draft

\title { Fluctuation Analysis of Human Electroencephalogram}

\author{Rudolph C. Hwa}
\address{Institute of Theoretical Science and Department of
Physics, University of Oregon, Eugene, OR 97403-5203}
\author{Thomas C. Ferree}
\address{
Electrical Geodesics, Inc., Riverfront
Research Park, Eugene, OR 97403\\
Computational Science Institute, University of Oregon,
Eugene, OR 97403}

\date{May 2001}
\maketitle
\begin{abstract} The scaling behaviors of the human
electroencephalogram (EEG) time series are studied using
detrended fluctuation analysis. Two scaling regions are found in
nearly every channel for  all subjects examined. The scatter
plot of the scaling  exponents for all channels (up to 129)
reveals the complicated  structure of a subject's brain
activity.  Moment analyses  are performed to extract the gross
features of all the  scaling exponents, and another universal
scaling behavior  is identified. A one-parameter description is
found to  characterize the fluctuation properties of the
nonlinear  behaviors of the brain dynamics.
\end{abstract}


In the study of electrical activities of the brain recorded by
electroencephalogram (EEG) various methods have been used to
extract different aspects of the neuronal dynamics from the
scalp potentials.  They range from the traditional linear
analysis that involves frequency decomposition, topographic
mapping, etc. \cite{eeg,nunez}, to time-frequency analysis  that
uses wavelet transform\cite{bla}, to nonlinear analysis  that is
particularly suitable for learning about the chaotic  behavior
of the brain\cite{chaos} or for quantifying physiological
conditions in nonlinear-dynamics terms \cite{jan,leh}. In this
paper we discuss the scaling behavior of the fluctuations  in
EEG in nonlinear analysis and show the existence of new
features in brain dynamics hitherto unrecognized. Moreover, we
propose a global measure of the spatio-temporal signals that has
potential utility in clinical and cognitive diagnostics.

Apart from being more suitable to analyze non-stationary time
series, the study of scaling behavior emphasizes the
relationship across time scales and provides a different
description of the time series than the conventional Fourier
power spectrum.  Conveniently it also liberates our result  from
the dependence on the magnitude of the voltage recorded by each
probe.  We aim to find what is universal among all  channels as
well as what varies among them. The former is  obviously
important by virtue of its universality for a given  subject;
how that universal quantity varies from subject to  subject
would clearly be interesting.  The latter, which is  a measure
that varies from channel to channel, is perhaps  even more
interesting, since that quantity has brain anatomical correlates
once the scalp potentials have been deconvolved to the cortical
surface.

Our procedure is to focus initially on one channel at a time.
Thus it is a study of the local temporal behavior and the
determination of a few parameters (scaling exponents) that
effectively summarize the fluctuation properties of the time
series. The second phase of our procedure is to describe the
global behavior of all channels and to arrive at one number that
summarizes the variability of these temporal measures across the
entire scalp surface.  This dramatic data reduction  necessarily
trades detail for succinctness, but such reduction  is exactly
what is needed to allow easy discrimination between  brain
states.

The specific method we use in the first phase is detrended
fluctuation analysis (DFA). This analysis is not new. It was
proposed for the investigation of correlation properties in
non-stationary time series and applied to the studies of
heartbeat \cite{peng} and DNA nucleotides \cite{pbh}.  It has
also been applied to EEG\cite{wat}, but with somewhat different
emphases than those presented here.  Since the  analysis
considers only the fluctuations from the local  linear trends,
it is insensitive to spurious correlations  introduced by
non-stationary external trends.  By examining  the scaling
behavior one can learn about the nature of  short-range and
long-range correlations, which are a  salient aspect of the
brain dynamics from the viewpoint  of complex systems theory.

Let an EEG time series be denoted by $y(t)$, where $t$ is
discrete time ranging from 1 to $T$. Divide the entire range of
$t$ to be investigated into $B$ equal bins, discarding any
remainder, so that each bin has $k={\rm floor}(T/B)$ time points. 
Within each bin, labeled $b\ (b=1,\cdots, B)$, perform a least-square
fit of $y(t)$ by a straight line, $\overline y_b(t)$, i.e.,
$\overline  y_b(t)=$ Linear-fit$[y(t)]$ for $(b-1)k<t\leq bk$.
That is the semi-local trend for the $\it b$th bin. Combine
$\overline y_b(t)$  for all $B$ bins and denote the $B$ straight
segments by
\begin{equation}
\overline y(k,t)=\sum_{b=1}^B \overline
y_b(t)\,\theta(t-(b-1)k)\,\theta(bk-t) \label{(1)}
\end{equation} 
for $1\leq t\leq kB$. Define
\begin{equation} F^2(k)={1\over {kB}}\sum_{t=1}^{kB}\
[\,y(t)-\overline y(k, t)\,]^2.
\label{(2)}
\end{equation}
$F(k)$ is then the RMS fluctuation from the semi-local trends in
$B$ bins each having $k$ time points, and is also a measure of
the fluctuation in each bin averaged over $B$ bins. The study of
the dependence of $F(k)$ on the bin size $k$ is the essence of
DFA \cite{peng,pbh}. If it is a power-law behavior
\begin{equation} F(k) \propto k^{\alpha} ,  \label{(3)}
\end{equation} then the scaling exponent $\alpha$ is an
indicator of the power-law correlations of the fluctuations in
EEG, and is independent of the magnitude of $y(t)$ or any
spurious trend externally introduced.

Resting EEG data were collected for six subjects using a
128-channel commercial EEG system, with scalp-electrode
impedences ranging from 10 to 40 k$\Omega$. The acquisition rate
is 250 points/sec with hardware filtering set between 0.1 and
100 Hz. After acquisition, $T\approx 10 $s lengths of
simultaneous time series in all channels are chosen, free of
artifacts such as eye blink and head movements.  At each time
point, the data are re-referenced to the average over all
electrodes.  This approximates the potentials relative to
infinity, and provides a more interpretable measure of  local
brain activity\cite{nunez}.  We investigate the range of $k$ 
from 3 to 500 in approximately equal steps  of $\ln k$.

% figure 1
\begin{figure}[ht]
\center\epsfig{figure=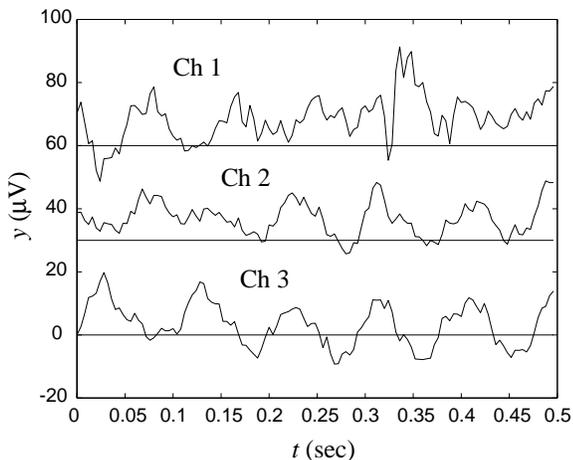,width=3.25in}
\caption{A sample of EEG time series in three channels. The
vertical scales of Ch 1 and Ch 2 are shifted upward by 60 and
30 $\mu$V, respectively.}
\end{figure}

In Fig.\,1 we show three typical time series $y(t)$ in three
widely separated channels for subject A, labeled 1-3, for
brevity.  While it is clear that both channels 2 and 3
have substantial 10 Hz  oscillations after 0.2 s, it is much
less apparent that there exist any scaling behaviors in all three
channels.  The corresponding values of $F(k)$ are shown in  the
log-log plot in Fig.\,2.  Evidently, the striking feature  is
that there are two scaling regions with a discernible  bend when
the two slopes in the two regions are distinctly  different.
With rare exceptions this feature is found in  all channels for
all subjects.  Admittedly, the extents of  the scaling regions
are not wide, so the behavior does not  meet the qualification
for scaling in large critical systems  or in fractal geometrical
objects. However, since the behavior  is so universal across
channels and subjects, it is a feature  of EEG that conveys an
important property of the brain activity  and should not be
ignored.

% figure 2
\begin{figure}[ht]
\center\epsfig{figure=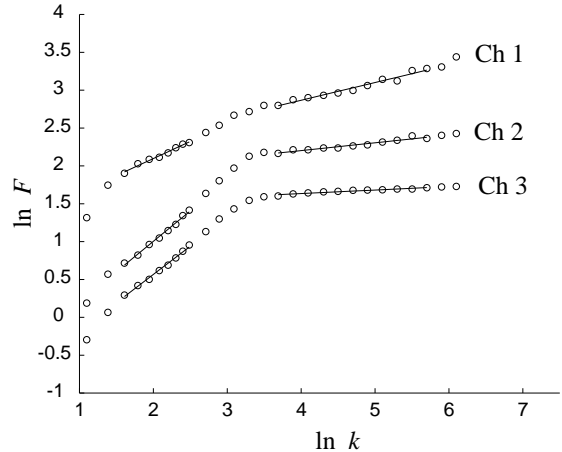,width=3.25in}
\caption{$F(k)$ vs $k$ for the three channels in Fig.\,1. The
vertical scales of Ch 1 and Ch 2 are shifted upwards by 1.0 and
0.5 units, respectively.}
\end{figure}

To quantify the scaling behavior, we perform a linear fit in
Region I for $1<\ln k < 2.5$ and denote the slope by
$\alpha_1$, and similarly in Region II for $3.5 < \ln k < 5.75$ with
slope denoted by $\alpha_2$.   Knowing the two straight  lines
in each channel allows us to determine the location of  their
intercept, $\ln\kappa$, which we define as the position  of the
bend in  $\ln k$. We find that, whereas $\alpha_1$  and
$\alpha_2$ can fluctuate widely from channel to channel,
$\kappa$  is limited to a narrow range in most subjects. The
average values of $\ln \kappa$ for  the six subjects are 3.45,
3.66, 3.10, 2.92, 2.60, and 3.12.

Since the only quantity in our analysis that has a scale is
$k$, the size of the bin on the time axis in which fluctuations
from the semi-local trend are calculated, the specific scale
$\kappa$ must correspond roughly to a particular frequency
$f$. If the data acquisition rate is denoted by $r$, then
$f=r/\kappa$. For $r=250$ points/sec, and $\ln\kappa=3.45$, 
we get $f=7.94$ Hz. That is near the midpoint separating the 
traditional $\alpha$ (8-13 Hz) and $\theta$ (5-8 Hz) EEG 
frequency bands. Region I is then at higher frequencies, 
Region II lower. Thus in this nonlinear analysis we have 
found the existence of a  specific frequency in every channel 
that separates different scaling behaviors.  A change in 
scaling exponent in physical systems is often attributable 
to distinct dynamical processes underlying the generation 
of the time series.  An interesting question is how this 
finding relates to EEG neurophysiology.

We now exhibit the values of $\alpha_1$ and $\alpha_2$ of all
channels for subject A in a scatter plot in Fig.\ 3. They vary
in the ranges: $0.19<\alpha_1<1.44$ and $0.018<\alpha_2<0.489$.
Whereas $\alpha_1$ is widely distributed, $\alpha_2$ is sharply
peaked at 0.1 and has a long tail. The value of
$\alpha=0.5$ corresponds to random walk with no correlation
among the various time points. For $\alpha\neq 0.5$ there are
correlations: Region I corresponds to short-range correlation,
Region II long-range, with $\kappa$ giving a quantitative
demarkation between the two. It is natural to conjecture that
high deviations of the $\alpha$ values from the average could 
be caused by pathological conditions.  Since the two scaling 
regions have corresponding frequency bands, one would likely 
want to focus on Region II to study brain states characterized 
by marked $\delta$ (1-4 Hz) and $\theta$ (5-8 Hz) activity.  
Sleep and cerebral ischemic stroke are two such examples.

% figure 3
\begin{figure}[ht]
\center\epsfig{figure=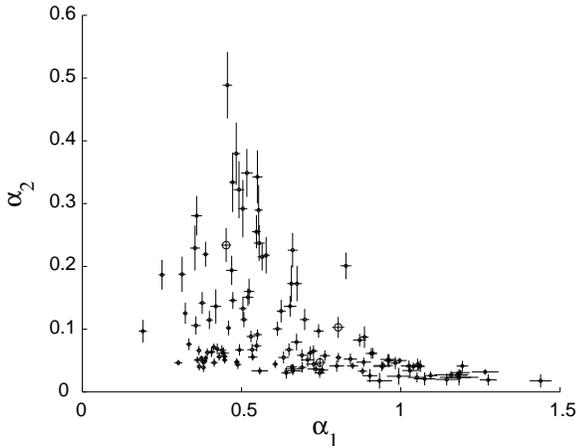,width=3.25in}
\caption{Scatter plot of $\alpha_2$ vs $\alpha_1$ for subject A.
The three channels exhibited in Figs.\,1 and 2 are shown as
big circles.}
\end{figure}

So far our consideration has focused on the temporal properties
of the time series in each channel. The scatter plot of
$\alpha_2$ vs $\alpha_1$ provides a view of the complicated
activities in all channels over the entire scalp. While the
detailed spatial structure may be of interest for relating
the values of $\alpha_i$ to brain anatomy, there is also a need 
for a general, overall description of all the pairs $\alpha_i$.  
A global measure for each subject could be of great use to 
specialists and non-specialists alike.  To that end, we now 
consider the relationship among the $\alpha_i$ values, or 
more precisely, the fluctuation of $\alpha_i$.

Let $x$ be either $\alpha_1$ or $\alpha_2$, and $N$ be the total
number of channels whose $\alpha_i$ values are under
consideration. Since no value of $\alpha_i$ has been found to
exceed 1.5 in the subjects we have examined, we consider the 
interval $0\leq x\leq 1.5$. Divide that interval into $M$ 
equal cells, which for definiteness we take to be $M=150$ 
here. Let the cells be labeled by $m=1,\cdots, M$, each having 
the size $\delta x=1.5/M$. Denote the number of channels whose 
$x$ values are in the $m$th cell by $n_m$. Define
\begin{equation} P_m = n_m\,/\,N.  \label{4}
\end{equation} It is the fraction of channels whose $x$ values
are in the range $(m-1)\,\delta x \leq x < m\,\delta x$. By
definition, we have
$\sum_{m=1}^M\,P_m=1$. In Fig.\,4 we show as an illustration the
two graphs of $P_m$ for subject A, three of whose EEG times
series are shown in Fig.\,1. The two graphs correspond to
$\alpha_1$ and $\alpha_2$, and are, in essence, the projections
of the scatter plot in Fig.\,3 onto the $\alpha_1$ and
$\alpha_2$ axes.

% figure 4
\begin{figure}[ht]
\center\epsfig{figure=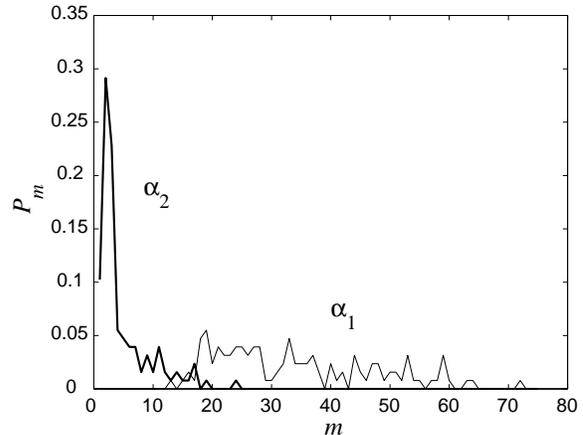,width=3.25in}
\caption{The distributions $P_m$ for $\alpha_1$ and
$\alpha_2$ for subject A. The bin size in $\alpha$ 
for this plot is 0.02.}
\end{figure}

Instead of studying the complicated structures of the
distributions $P_m$ themselves, it is more convenient to examine
the moments of $P_m$. Thus we define the normalized moments
\begin{equation} G_q=\sum_{m=1}^M\,m^q\,P_m\left/
(\sum_{m=1}^M\,m\,P_m)^q,\right.
\label{5}
\end{equation} where $q$ is a positive integer, although it can
be a continuous variable. Since $x$ can be either $\alpha_1$ or
$\alpha_2$, we shall use
$P_m^{(1)}$ and $P_m^{(2)}$ to denote the two distributions, and
$G_q^{(1)}$ and $G_q^{(2)}$, respectively, for their moments.
Since these moments are averages of $(m/\overline m)^q$, where
$\overline m$ is the average-$m$, they are not very sensitive to
$\overline m$ itself. They contain the essence of the
fluctuation properties of $\alpha_{1,2}$ in all channels.

% figure 5
\begin{figure}[ht]
\center\epsfig{figure=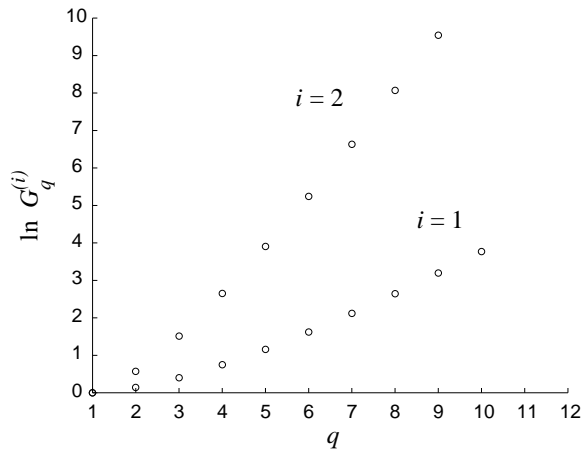,width=3.25in}
\caption{The $q$ dependence of $\ln G_q^{(i)}$ for subject A.}
\end{figure}

In principle, it is possible to examine also the moments for
$q<0$, which would reveal the properties of $P_m$ at low values
of $m$. However, the accuracy of our data is not too reliable 
for low-$k$ analysis, since the 60 Hz noise due to ambient 
electric and magnetic fields has not been cleanly filtered out. 
In this paper, therefore, we restrict our study to only the 
positive $q$ values.  For high $q$, the large $m/\overline m$ 
parts of $P_m^{(1,2)}$ dominate $G_q^{(1,2)}$.

In Fig.\,5 the $q$-dependences of $\ln G_q^{(1,2)}$ are shown
for the distributions exhibited in Fig.\,4 for
$0\leq q\leq 10$. They appear to depend on $q$ quadratically.
The relationship between $G_q^{(1)}$ and
$G_q^{(2)}$ is, however, extremely simple, as can be shown by
plotting them against each other in a log-log graph.  In Fig.\,6
we show the straightline behavior for three of the six subjects,
all of whose EEG time series have been analyzed in the same
method described here. Thus we can associate a slope $\eta$ to
every subject, and conclude that there exists a universal scaling
behavior
\begin{equation} G_q^{(2)} \propto
\left(\,G_q^{(1)}\,\right)^{\eta}.
\label{6}
\end{equation} This remarkable behavior is valid for all
subjects examined, but the exponent $\eta$ varies from subject
to subject. Thus we have discovered a measure that characterizes
all the $\alpha_i$ values of a subject.

% figure 6
\begin{figure}[ht]
\center\epsfig{figure=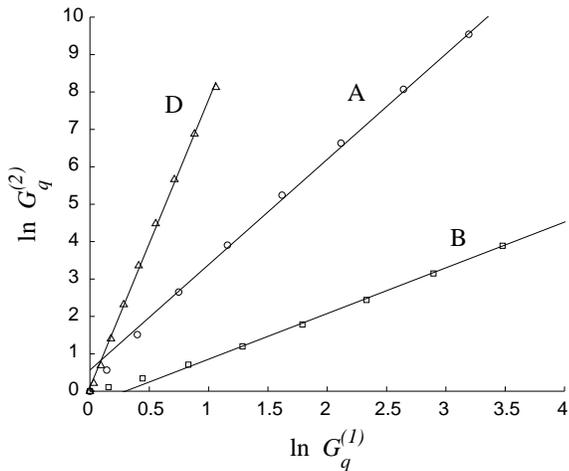,width=3.25in}
\caption{A log-log plot of $G_q^{(2)}$ vs $G_q^{(1)}$ for
three subjects. The solid lines are linear fits of the large $q$
parts of the plots.}
\end{figure}

We label the six subjects whom we have examined as A-F. Their
$\eta$ values are given below behind the subject labels.

\begin{center}
A\,(2.82),  B\,(1.22),  C\,(1.74),  D\,(7.68),  E\,(6.39),
F\,(3.29)

\end{center}

At this stage of our investigation we have not yet arrived at a
point where we can give a definitive correlation between the
values of $\eta$ and some specific aspect of the brain activity.
It is, nevertheless, of interest to note that the subjects A-C
were regarded as healthy control subjects, while  D-F have each
had recent encounter with ischemic stroke.

Before any tentative correlations may be inferred, we emphasize 
the need for caution since many factors were involved in data 
acquisition, e.g., varying degrees of awakeness of the subjects.   
Clinical and cognitive application is not the main object of 
this paper.  However, the intriguing possibilities of the method 
presented here suggest extensive application of the analysis to 
many more subjects and for further investigation on what the 
scaling behaviors of the spatio-temporal EEG reveal.  Such 
behaviors can undoubtedly be useful in guiding future theoretical 
development.

To summarize, we have studied the scaling properties of the
fluctuations in the time series, and found a number of
features previously unrecognized in power spectrum analysis. The
characterization of the temporal behavior of each channel by
just two parameters, $\alpha_1$ and $\alpha_2$, has led  to the
discovery of the $\eta$ exponent.  Although $\eta$ is  obtained
by considering the global property of all channels, we envisage
its extension to the study of local properties by the
application of the moment analysis to partitioned regions of the
scalp. Future possibilities of this approach to the
investigation of EEG signals seem bountiful.

We are grateful to Dr.~Phan Luu and Prof.~Don Tucker for supplying
the EEG data for our analysis.  This work was supported, in
part,  by the U.\,S.\,Department of Energy under Grant No.\,
DE-FG03-96ER40972, and the National Institutes of Health under
Grant No.\,R44-NS-38829.

\end{document}